\documentstyle[12pt]{article}

\def\be{\begin{equation}}
\def\ee{\end{equation}}
\def\ben{$$}
\def\een{$$}
\def\ba{\begin{array}{c}}

\def\ea{\end{array}}
\def\p{\partial}
\begin{document}

\titlepage
\begin{center}
.

\vspace{2cm}

{\Large \bf Short-range oscillators in power-series picture
 }\end{center}

\vspace{10mm}

\begin{center}
Miloslav Znojil

\vspace{3mm}

odd\v{e}len\'{\i} teoretick\'{e} fyziky,\\ \'{U}stav jadern\'e
fyziky AV \v{C}R, 250 68 \v{R}e\v{z}, Czech Republic\footnote{
\today; e-mail: znojil@ujf.cas.cz}\\

\end{center}

\vspace{5mm}

\section*{Abstract}

The class of short-range potentials $V^{[M]}(x) =\sum_{m=2}^{M}
(f_m+g_m\sinh x)/\cosh^m x$ is considered as an asymptotically
vanishing phenomenological alternative to the popular anharmonic
long-range $V(x)=\sum_{n=2}^{N}h_n x^n$. We propose a method which
parallels the analytic Hill-Taylor description of anharmonic
oscillators and represents all the wave functions $\psi^{[M]}(x)$
non-numerically, in terms of certain infinite hypergeometric-like
series. In this way the well known exact $M=2$ solution is
generalized to any $M> 2$.

\vspace{10mm}

PACS
\hspace{5mm} 03.65.Ge
\hspace{5mm} 02.30.Gp
\hspace{5mm} 02.30.Hq
\hspace{5mm}
03.65.Db

\newpage

\section{Introduction}

A routine numerical solution of an {\em asymmetric}
Schr\"{o}dinger bound-state problem on the line $x \in
(-\infty,\infty)$ requires a careful verification \cite{Asym}. One
needs non-numerical asymmetric models. For this purpose we may use
the shifted harmonic oscillator, Morse's well and the two
scarf-shaped hyperbolic forces. All of these models (cf. Table
\ref{shapeinp}) are listed in review \cite{Khare} as possessing
the {\em complete} solution in closed form.

There exist {\em incompletely} solvable polynomials $ V(x)= a\,x +
b\,x^2+ \ldots + z\,x^N$ \cite{Magyari} and multi-exponentials
$V(x)= a\,e^{-x} + b\,e^{-2x}+ \ldots + z\,e^{-Nx}$ \cite{Cizek}.
They extend the possible tests and further non-numerical
applications beyond $N=2$. In a puzzling contrast, a natural
generalization
 \be
 V^{[M]}(x) = \sum_{m=2}^{M}  \frac{f_m}{\cosh^{m} x}
 + \sinh x \,\sum_{n=1}^{M}  \frac{g_n}{\cosh^{n} x}
  \label{RMG}
 \ee
of the remaining two items in Table \ref{shapeinp} is not amenable
to the similar elementary treatment \cite{Ushveridze}. This
distracts attention from the hyperbolic oscillators (\ref{RMG}) in
spite of their obvious phenomenological as well as purely
mathematical appeal.

In the present paper we shall return to several formal as well as
descriptive parallels between the separate items in Table
\ref{shapeinp}. On their basis we shall propose and describe a new
semi-analytic approach to the ``neglected" family (\ref{RMG}).

In Section 2 we recall the harmonic and Morse oscillators and
their $N>2$ generalizations as our overall methodical guide. In
the language of the well known Lanczos method \cite{Wilkinson} we
underline the key role of simplicity of the repeated action of the
Hamiltonian upon a suitable trial state $|0\rangle$. An
appropriate choice of this initial ket vector is able to inspire
some of the existing non-numerical power series solutions. In this
setting the Lanczos approach is shown to find its natural
re-incarnations in the well known method of Hill determinants
\cite{Hillove} as well as in the symmetric Jost-solution method of
ref. \cite{LMP}.

In Section 3 we show that and how the latter two examples pave the
way towards eq. (\ref{RMG}) with any $M \geq 2$. In a full
parallel to the polynomial case we construct the asymptoticaly
correct bound state solutions which all retain a recurrently
defined power-series structure. Via an appropriate $D-$dimensional
partitioning of the basis we preserve their connection to the two
remaining exactly solvable hyperbolic $M=2$ examples of
Table~\ref{shapeinp}.

Section 4 illustrates the technical details at the first
nontrivial $D=2$. We contemplate there a spatially anti-symmetric
$M=2$ exercise (\ref{RMG}) using $f_2=g_1=0$. We detail the proof
of the point-wise convergence of our ``partitioned hypergeometric"
wave functions. We show how the symmetry considerations
significantly simplify the construction and matching of our wave
functions near the origin.

Section 5 adds a short summary.

\section{The method}

\subsection{Wave functions in the Lanczos basis}

The Lanczos numerical eigenvalue method \cite{Wilkinson} works
with a set $\{\,|n\rangle\,\}$ of the basis ket vectors which are
generated via a repeated action of the Hamiltonian $H$ upon an
initial vector $|0\rangle$. In a slight generalization of this
procedure one has to assume that the action of the full
Schr\"{o}dinger operator $H-z$ upon each ket $|n\rangle$ may be
represented as a linear superposition over the same set of the
kets \cite{jasam},
 \be
  (H-z)|n\rangle= |0\rangle \cdot Q_{0,n}(z) + |1\rangle
\cdot Q_{1,n}(z) + \ldots\ .
 \label{ql}
 \ee
With a matrix of functions $Q_{m,n}(z)$ (cf. \cite{Lanczos}, p.
257) we may abbreviate
 \ben
  \left[ \ba (H-z)|0\rangle,  (H-z) |1\rangle,
\ldots
\\
\ea \right]
 \equiv
 (H-z)\ \left[ \ba |0\rangle,  |1\rangle,  |2\rangle,
\ldots
\\
\ea \right]
 \equiv
 (H-z)\ \left| \ba
 X
\\
\ea \right \} \ ,
 \een
 \ben
 (H-z)\ \left| \ba
 X
\\
\ea \right \}
=
 \left| \ba
 X
\\
\ea \right \}
 \cdot Q(z)\
  \een
and solve any linear homogeneous equation $(H-E)|y\rangle = 0$ by
the ansatz
 \be
|y\rangle =
\sum_{n=0}^\infty \, |n\rangle\, h_n
 \equiv
 \left | X \right \} \ \vec{h}
\ .
  \label{sol}
 \ee
Provided that the separate lanczosean kets are linearly
independent the resulting identity $|X\} Q(z) \vec{h}=0$ may be
interpreted as a system of conditions
 \be
Q(z) \vec{h} =0\
 \label{uni}
 \ee
The practical applicability of this recipe relies upon several
tacit assumptions. Most often one chooses the set
$\{\,|n\rangle\,\}$ as a common harmonic oscillator basis
\cite{Chaudhuri}. It is orthonormal ($\{ X | X \} = I$) and
complete ($|X \}\,\{ X|=Id$) and we may truncate the linear set
(\ref{uni}) to the mere routine matrix diagonalization
 \be
 \sum_{n=0}^{\cal M} \left [Q(0)-E\,I\right ]_{m,n} {h}_n =0, \ \ \ \ \ \ \
 m=0, 1, \ldots, {\cal M}, \ \ \ {\cal M} \gg 1\ .
 \label{matri}
 \ee
This is a textbook variational recipe and its secular equation
 \be
\det Q(E)=0
\label{Hills}
 \ee
determines the spectrum numerically \cite{Hioe}.

A non-variational and less numerical modification of the
construction may be based on a more sophisticated choice of the
Lanczos basis. Various linear algebraic algorithms of such a type
are used to solve various Schr\"{o}dinger equations in
applications \cite{Whitehead}. Let us recall two examples as our
methodical guide.

\subsection{Anharmonic example}

Both the above-mentioned multi-exponential and polynomial
oscillators prove mutually equivalent after a change of variables
\cite{Sukhatme}. Their ``canonical" \cite{Classif} representation
 \be
V(x)= \frac{g_{-1}}{r^2} + g_1r^2 + g_2r^4+\ldots +
g_{2N-1}r^{4N-2} , \ \ \ \ \ r \in (0, \infty)
 \label{SEpol}
 \ee
is easily tractable by the variational algorithms. In the less
numerical power-series approaches \cite{Hille} the harmonic kets
are being replaced by their mere power-law components $\langle r |
n \rangle = \langle r |0^{[HO]} \rangle \cdot r^n$. This leads to
an asymmetric matrix $Q(z)$. Its linear algebraic eq. (\ref{uni})
proves often solvable as a very simple recurrent specification of
the coefficients $h_n$ in eq. (\ref{sol}) (cf. ref. \cite{Drozdov}
for more details).

An even more ambitious reduction of $Q$ may be achieved after an
anharmonic choice of the initial $|0\rangle $. According to
Magyari \cite{Magyari} this assigns a few elementary bound-state
solutions to many multi-exponential and polynomial potentials at
certain exceptional couplings. At {\em arbitrary} couplings and
energies the same option $|0\rangle $ may provide an extremely
compact infinite-dimensional algebraic secular equation
(\ref{uni}). For illustration let us consider the famous sextic
oscillator example of ref. \cite{Singh}. With $N=2$ in eq.
(\ref{SEpol}), denoting $g_3=16\,\alpha^2$ and $g_2= 16\,\alpha\,
\beta$ and using the WKB-inspired postulate
 \be
\langle r | n \rangle =  r^{n+\ell+1} e^{-\alpha\,r^4 -
\beta\,r^2}\ , \ \ \ \ \ \alpha > 0\
  \label{hilc}
 \ee
we get the tridiagonal quasi-Hamiltonian
 \be
 Q(E)= \left(
\begin{array}{ccccc} \alpha_0&\gamma_1&0&0&\ldots\\
\beta_0&\alpha_1&\gamma_2&0&\ldots\\
0&\beta_1&\alpha_2&\gamma_3&\ldots\\ &&\ddots&\ddots& \ea \right)\
. \label{con}
 \ee
Its equation (\ref{uni}) may safely be interpreted as an
infinite-dimensional limit of the truncated diagonalization
(\ref{matri}) provided only that $g_2> 0$ \cite{jadva}. The three
nonzero diagonals in eq. (\ref{con}) have to be compared with the
seven-diagonal structure of the Hamiltonian in the usual
orthogonalized harmonic oscillator basis.

For $g_2 \leq 0$ and at a special discrete set of the couplings
$g_1$ the infinite-dimensional tridiagonal secular Hill
determinant factorizes and the recipe reproduces {\em a part} of
the spectrum correctly \cite{Singh}. In all the other cases the
WKB-compatible Lanczos basis ceases to be adequate. The
Hill-determinant recipe (\ref{Hills}) loses its relation to the
correct asymptotic boundary conditions and the basis (\ref{hilc})
must be regularized for certain hidden-symmetry reasons
\cite{Hautot}. More diagonals necessarily appear in eq.
(\ref{con}). Otherwise, one gets wrong results from the truncated
eq. (\ref{matri}) even in its infinite-dimensional limit
\cite{Tater}.

Virtually no similar constructions of our short-range hyperbolic
oscillators seem to appear in the current literature. In the
present paper we intend to explain the difference and develop a
new semi-analytic approach to eq. (\ref{RMG}). Our construction
will fairly closely parallel the formalism of the Hill-determinant
method. In our second preparatory step the appropriately modified
choice of the Lanczos basis will be illustrated via the
symmetrized Rosen-Morse or scarf model of Table~\ref{shapeinp}.

\subsection{P\"{o}schl-Teller example }

Formula (\ref{RMG}) with $M=2$, attraction $f =
-\lambda(\lambda-1)$ {\em and} vanishing $g=0$ defines the
bell-shaped and spatially symmetric P\"{o}schl-Teller well $
V^{(PT)}(x)= {f}/{ \cosh^{2} x}$ \cite{Poeschl}. The functional
form of the optimal lanczosean kets is more or less uniquely
deduced, very much in the spirit of the ``most ambitious" WKB-like
choice in eq. (\ref{hilc}) above, from the available exact
solutions,
 \be
\langle x | n\rangle = \xi_{n,p,q,\kappa}(x)={\sinh^{1-q} x \over
 \cosh^{\kappa+2n+p} x}\ \in\ L_2(-\infty,\infty)\ .
 \label{party}
  \label{fullie}
   \ee
All these basis states possess the even or odd parity at $q=1$ or
$q=0$, respectively. Within this subsection let us fix $p \equiv
1-q$. Then, the action of the full Hamiltonian
$H^{(PT)}=-\p_x^2+V^{(PT)}(x)$ on our symmetrized/anti-symmetrized
states (\ref{party}) becomes particularly transparent. For
energies $E = -\kappa^2$, it is characterized by the mere
two-diagonal matrix
 \be Q(E)= \left(
\begin{array}{ccccc} \alpha_0&0&0&0&\ldots\\
\beta_0&\alpha_1&0&0&\ldots\\ 0&\beta_1&\alpha_2&0&\ldots\\
&&\ddots&\ddots& \ea \right)\ \label{condia}
 \ee
with the vanishing uppermost element $\alpha_0=0$. The bound-state
solutions (\ref{sol}) of our Schr\"{o}dinger differential equation
read
 \be \frac{1}{h_0}\,\langle x | y\rangle = |0\rangle - |1\rangle
\cdot \frac{\beta_0}{\alpha_1} + |2\rangle \cdot
\frac{\beta_0\beta_1}{\alpha_1\alpha_2} +\ldots\ .
 \label{series}
 \ee
As long as they are defined by the really elementary two-term
recurrences (\ref{sol}),
 \be
\left(
\begin{array}{cccc}
0&0&0&\ldots\\ f+(\kappa+p)(\kappa+p+1),& -4(\kappa+1)& 0&\ldots\\
0&f+(\kappa+p+2)(\kappa+p+3),& -8(\kappa+2)& \ldots\\
\vdots&\ddots&\ddots&\ddots \ea \right) \left(
\begin{array}{c}
h_0\\ h_1\\ h_2\\ \vdots \ea \right)=0 \label{recu}
 \ee
our solution $|y\rangle$ coincides with the Gauss hypergeometric
series,
 \be
\langle x | y\rangle=h_0 \tanh^p x\,\frac{1}{\cosh^{\kappa}
x}\,_2F_1 \left (
 \frac{\kappa+p+\lambda}{2},
 \frac{\kappa+p+1-\lambda}{2};
 1+\kappa;\frac{1}{\cosh^2 x} \right )\ .
 \label{inser}
 \ee
It is defined on a half-axis, say, $x \geq 0$. Fortunately, due to
the manifest symmetry or anti-symmetry of the physical solutions
the necessary analytic continuation across the origin proves
equivalent to the termination of this infinite series. The well
known Jacobi polynomial solutions are obtained at each physical
energy \cite{LMP}.

\section{Partitioned expansions}

We may conclude that the description of bound states by the
infinite series (\ref{sol}) proves easy and efficient not only in
the Hill-determinant setting of section 2.2 but also in an
alternative Jost-solution spirit of section 2.3. We intend to
extend the parallelism far beyond the trivial example of section
2.3.

The action of the kinetic energy $T=-\p_x^2$ on the basis
(\ref{party}) conserves {\em both} the independent parity-like
parameters $p$ and $q$. The same conservation law is obeyed by the
single-term symmetric potentials $ V_s^{(M)}(x)= {f}/{ \cosh^{M}
x} $ with the even exponents $M=2K$. The rule is broken by the
general Hamiltonians containing superpositions (\ref{RMG}) of the
symmetric and anti-symmetric components $V_s^{(M)}(x)$ and
$V_a^{(N)}(x)= g\,{\sinh x}/{ \cosh^{N} x}$, respectively.
Nevertheless, the full basis (\ref{party}) numbered by a composite
index $\mu =\mu(n,p,q)= 4n+2p+q \geq 1$ (as $\xi_{n,p,q,\kappa}(x)
\equiv \langle x | \Xi_\mu \rangle $, $\mu = 1, 2, \ldots$) proves
reducible for all the single-term potentials $V_{s,a}^{(N)}(x) =
\pm V_{s,a}^{(N)}(-x)$ of a definite parity.

\subsection{Symmetric potentials $V(x)=V(-x)$}

We may choose the initial Lanczos ket $|0\rangle$ either as the
spatially symmetric (and asymptotically correct) element $\langle
x |\Xi_{\mu(0,0,1) }\rangle \equiv \cosh^{-\kappa} x$ with $p=0$
and $q=1$ {\em or} as its anti-symmetric analogue $\langle x
|\Xi_{\mu(0,1,0) }\rangle \equiv {\sinh x\cdot \cosh^{-\kappa-1} x
}$ with $p=1$ and $q=0$. In both these cases, all the Hamiltonian
operators $T+V_s^{(2K)}(x)$ become compatible with recurrences
(\ref{ql}) in the two alternative bases
 \ben |0\rangle, \, |1\rangle, \,
|2\rangle,\, \ldots = | \Xi_{\mu(0,0,1) }\rangle , |
\Xi_{\mu(1,0,1) }\rangle , | \Xi_{\mu(2,0,1) }\rangle \ldots
 \equiv | \Xi_1 \rangle , |\Xi_5 \rangle , | \Xi_9 \rangle ,
\ldots\ , \een \ben |0\rangle, \, |1\rangle, \, |2\rangle, \,
\ldots  = | \Xi_{\mu(0,1,0) }\rangle , | \Xi_{\mu(1,1,0) }\rangle
, | \Xi_{\mu(2,1,0) }\rangle \ldots \equiv | \Xi_2 \rangle , |
\Xi_6 \rangle , | \Xi_{10} \rangle , \ldots \
 \een
with $ p = 1-q = 0$ or $1$, respectively.  After we abbreviate
$a_j=-j\,(2\kappa+j)$ and $b_j=(\kappa+j)(\kappa+j+1) $, this
enables us to reproduce the two-diagonal P\"{o}schl-Teller
realization of $Q=Q^{(p)}$ at $K=1$,
 \ben Q^{(0)} = \left(
\begin{array}{cccc}
0&0&0&\ldots\\ f+b_0,& a_2& 0&\ldots\\ 0&f+b_2,&a_4&
\\
\vdots &&\ddots&\ddots \ea \right), \  \ \ \ Q^{(1)} = \left(
\begin{array}{cccc}
0&0&0&\ldots\\ f+b_1,& a_2& 0&\ldots\\ 0&f+b_3,&a_4&
\\
\vdots &&\ddots&\ddots \ea \right).
 \een
In the ``first unsolvable" case with $K=2$ the coupling $f$ moves
one step down,
 \ben Q^{(0)} = \left(
\begin{array}{c|cc|cc}
0&0&0&0&\ldots\\ \hline
 b_0,& a_2& 0&0&\ldots\\ f&b_2,&a_4&0&\ldots\\
\hline 0&f&b_4,&a_6&\\ \vdots&&\ddots&\ddots&\ddots \ea \right), \
\ \  \ \ \ Q^{(1)} = \left(
\begin{array}{c|cc|cc}
0&0&0&0&\ldots\\ \hline b_1,& a_2& 0&0&\ldots\\
f&b_3,&a_4&0&\ldots\\ \hline 0&f&b_5,&a_6&\\
\vdots&&\ddots&\ddots&\ddots \ea \right)\ .
 \een
Partitioning indicated by the auxiliary lines tries to preserve
the same two-diagonal pattern as above. At $K=3$ we have,
similarly,
 \ben
Q^{(0)} = \left(
\begin{array}{c|ccc|c}
0&0&0&0&\ldots\\ \hline
 b_0,& a_2& 0&0&\ldots\\ 0&b_2,&a_4&0&\ldots\\
f&0&b_4,&a_6&\\ \hline &\ddots&&\ddots&\ddots \ea \right), \ \ \ \
\ \ Q^{(1)} = \left(
\begin{array}{c|ccc|c}
0&0&0&0&\ldots\\ \hline b_1,& a_2& 0&0&\ldots\\
0&b_3,&a_4&0&\ldots\\ f&0&b_5,&a_6&\\ \hline
&\ddots&&\ddots&\ddots \ea \right)
 \een
and so on. The dimension of partitions grows linearly with $M=2K$
as $D=K$.

The second series of the symmetric potentials $V_s^{(2K+1)}(x) =f/
\cosh^{ 2K+1} x$ with the {\em odd} powers $M=2K+1$ must be
investigated separately. It acts on our parity-preserving basis in
such a way that the conservation of the quantum number $p$ is
broken. The following $q-$preserving bases must be used,
 \ben |0\rangle,
\,|1\rangle,  \, |2\rangle,\,\ldots \equiv  |\Xi_2\rangle,
|\Xi_4\rangle, |\Xi_6\rangle, |\Xi_8\rangle, \ldots, \ \ \ \ \ \ \
q = 0,
 \een
 \ben
 |0\rangle, \,|1\rangle,  \, |2\rangle,\,\ldots
\equiv   |\Xi_1\rangle, |\Xi_3\rangle, |\Xi_5\rangle,
|\Xi_7\rangle, \ldots, \ \ \ \ \ \ \ q = 1.
 \een
At $K=0$ the new lower triangular matrices $Q =Q^{[q]}$ contain
just the three nonzero neighboring diagonals.  For a preservation
of the two-diagonal denotation it is sufficient to switch to the
$D=2$ partitioning.  Similarly, a three-dimensional partitioning
is needed at $K=1$. With the further increase of $K$ the dimension
$D=2K+1$ grows more quickly.

\subsection{Anti-symmetric potentials $V(x)=-V(-x)$}


The class of the anti-symmetric forces $V_a^{(2L)}(x)$ with $L\geq
1$ inter-relates the basis states with different parities $q$. The
value of the index $p$ is conserved,
 \ben V_a^{(2L)}(x)
|\Xi_{\mu(n,p,q)}\rangle=(1-q)\,g\cdot
|\Xi_{\mu(n+L-1,p,1-q)}\rangle +(-1)^{1-q} g\cdot
|\Xi_{\mu(n+L,p,1-q)}\rangle .
 \een
The Hamiltonian $T+V^{(2L)}_a$ acts transitively on the following
two reduced Lanczos bases,
 \be
|0\rangle, \,|1\rangle,  \, |2\rangle,\ldots \equiv |\Xi_1\rangle;
|\Xi_4\rangle, |\Xi_5\rangle; |\Xi_8\rangle, |\Xi_9\rangle;
\ldots, \ \ \ \ \ \ \ p = 0 \label{asya} \ee
\be
 |0\rangle, \,|1\rangle, \ldots
\equiv  |\Xi_2\rangle, |\Xi_3\rangle; |\Xi_6\rangle,
|\Xi_7\rangle; |\Xi_{10}\rangle, |\Xi_{11}\rangle; \ldots, \ \ \ \
\ \ \ p = 1. \label{asyb}
 \ee
Marginally, we may note that at $L=0$ the structure of the matrix
$Q$ ceases to be triangular. This seems closely related to the
asymptotic asymmetry of the $g_1 \neq 0$ potentials
$V^{[M]}(-\infty)=-g_1 \neq V^{[M]}(\infty)=+g_1$ and to their
anomalous non-Jost solvability via a change of variables at $M=2$
(cf., e.g., \cite{MorFes}). In this subsection we shall assume
that $g_1 \equiv 0$, therefore. This constraint is further
supported by the observation that at $M=1$ the monotonic
$V_a^{(1)}(x)= g_1\cdot \tanh x $ itself cannot generate any bound
states at all. Thus, our study of the anti-symmetric models has to
start at the exactly solvable $V_a^{(2)}(x)= g\,{\sinh x}/{
\cosh^{2} x}$ (cf. Table \ref{shapeinp}).

This anti-symmetric scarf (AS) potential $V_a^{(2)}(x)\equiv
V^{[AS]}(x)$ is extremely suitable for methodical purposes. Its
significance is connected to the fact that our basis
(\ref{fullie}) is not tailored precisely to its exact solvability.
A $D=2$ partitioning is needed. In the reduced bases (\ref{asya})
and (\ref{asyb}) its recommended boundaries are marked by the
semi-colons. For all the $L = 2,3,\ldots$ descendants
$V_a^{(2L)}(x)$ of the AS example the size $D$ of partitions will
grow due to the downward shift of the constant $g$ again.


The action of the last class $V_a^{(2L+1)}(x)= g\cdot\sinh x \cdot
\cosh^{-2L-1} x$ of the simplified single-term potentials on the
kets (\ref{party}) looks irreducible.  The impression is wrong.
After we introduce a new quantum number $I \equiv 2p+q\ ({\rm
modulo}\ 4)$, the basis elements with $I = 0$ and $I = 3$ never
mix with their $I = 1$ and $I = 2$ counterparts. For both the
initial choices of $|0\rangle= |\Xi_1\rangle$ and
$|0\rangle=|\Xi_2\rangle$ we arrive at the same output, \ben
 |0\rangle,
\, |1\rangle,\, |2\rangle, \ldots \equiv |\Xi_{1\ or \ 2}\rangle,
|\Xi_5\rangle, |\Xi_6\rangle, |\Xi_9\rangle, |\Xi_{10}\rangle,
|\Xi_{13}\rangle, \ldots\ .
 \een
The difference between the two matrices $Q$ will only lie in their
elements.

\subsection{Asymmetric Lanczos kets }

Asymmetric oscillators (\ref{RMG}) admit a non-conservation of
parity by each Lanczos element $|n\rangle$ separately. The
functions
 \be
\langle x|n\rangle =\xi_{n,p,q,a,\kappa}(x) = {\sinh^{1-q} x \over
\cosh^{\kappa+2n+p} x} \,e^{a\,\arctan(\sinh x)}\ \in\
L_2(-\infty,\infty)
 \label{genepar}
 \ee
generalize their $a=0$ predecessors (\ref{party}) and represent a
very good new candidate since, due to the presence of a new
parameter $a$, the number of the new terms in eq. (\ref{ql}) may
be lowered, for any potential (\ref{RMG}), more efficiently. First
of all, this implies that we may admit the nonzero $g_1$ again.
Via a suitable choice of the value of $a$ we shall be able to
reproduce {\em all} the ``missing" (viz., Rosen Morse and scarf)
{terminating} solutions of ref. \cite{Khare} or
Table~\ref{shapeinp}.

At $a \neq 0$ also the action of an arbitrary hyperbolic
Hamiltonian remains transparent and elementary in the purely
kinetic limit,
 \ben
 {\xi''_{n,p,q,a,\kappa}(x)  \over \xi_{n,p,q,a,\kappa}(x) }=
(\sigma +q-1)^2 +{a^2-\sigma(\sigma+1) -(2\sigma+1)a\,\sinh x
\over \cosh^2 x} +(q-1)\,{ q-2a\,\sinh x \over \sinh^2 x}.
 \een
Here, $\sigma=\sigma(n,p)=\kappa+2n+p$ and the prime denotes the
differentiation with respect to $x$. The action of the purely
kinetic Hamiltonian $T = -\p^2_x$ on our innovated kets $\langle x
| \Xi_\mu \rangle \equiv \xi_{n,p,q, a,\kappa}(x)$ may employ the
multi-indices $\mu(n,p,q)= 4n+2p+q $ again,
 \ben T| \Xi_{\mu(n,p,0) }\rangle = -(\sigma-1)^2\,|
\Xi_{\mu(n,p,0) }\rangle +(2\sigma-1)\,a\,| \Xi_{\mu(n,p,1)
}\rangle + \een \ben +(\sigma^2+\sigma-a^2)\, | \Xi_{\mu(n+1,p,0)
}\rangle -(2\sigma+1)\,a\, | \Xi_{\mu(n+1,p,1) }\rangle,
 \een
 \ben
T| \Xi_{\mu(n,p,1) }\rangle = - \sigma^2\, | \Xi_{\mu(n,p,1)
}\rangle +  (2\sigma+1)\,a\, | \Xi_{\mu(n+1,p,0) }\rangle
+(\sigma^2+\sigma-a^2)\, | \Xi_{\mu(n+1,p,1) }\rangle\ .
 \een
The kinetic matrix elements of $Q$ depend on $\sigma$ and $a$ and
all of them increase with $n$. Due to the presence of the new
parameter $a$ the kinetic operator $T$ inter-twins the states
(\ref{genepar}) with different parities $q=0,\,1$.  The states
with different $p= 0,\,1$ stay decoupled.

\subsection{Partitioned hypergeometric-like series}

Our present proposal may be summarized as an application of
expansions (\ref{sol}) to potentials (\ref{RMG}) inspired by the
analogies between the P\"{o}schl-Teller and harmonic oscillators.
The feasibility of our construction stems from the fact that the
action of the present class of Hamiltonians on the suitable
Lanczos kets may be characterized by the lower triangular matrices
$Q(z)$. Their partitioning brings us back to the two-diagonal
pattern of eq. (\ref{condia}) and replaces its scalars $\alpha_j$
and $\beta_j$ by the respective two-dimensional submatrices $A_j$
and $B_j$,
 \be Q= \left(
\begin{array}{ccccc}
A_0&0&0&0&\ldots\\ B_0&A_1&0&0&\ldots\\ 0&B_1&A_2&0&\ldots\\
&&\ddots&\ddots& \ea \right) \ . \label{qls}
 \ee
In both the respective $a=0$ and $a\neq 0$ bases (\ref{party}) and
(\ref{genepar}) the $D-$plets of kets $\left ( |m+1\rangle,\,
|m+2\rangle, \ldots, |m+D\rangle \right ) $ with $m=m(n)=nD-d_0$
and with any $d_0$ may be denoted as $ ||n\rangle\rangle $. In
such an abbreviated notation our linear system (\ref{uni}) implies
the recurrence relations
 \be
  {F_n} \equiv \left( \ba h_{m(n)+1}\\ \ldots\\ h_{m(n)+D} \ea
\right) = -(A_n)^{-1} B_{n-1} F_{n-1},
 \ \ \ \  \ n =  1,2, \ldots\
 \label{reccc}
 \ee
which define the $D-$dimensional vectors of coefficients in terms
of finite products of the certain $D\times D-$dimensional
matrices. In place of $d_0=1$ in a consequently $D-$dimensional
``democratic" partitioning we may use the shift $d_0=D$. Both
these options appear in our AS example where we recommended
$d_0=D-p$. The latter one is globally preferable as it leaves the
uppermost element of $Q$ vanishing, $A_0=0$. The initial array
$F_0$ degenerates to the mere scalar norm then.

At any $d_0$ the formal solution (\ref{sol}) of the
Schr\"{o}dinger equation $(H-E)|y\rangle=0$ may be re-written in
the form of the double or partitioned sum,
 \be
|y\rangle = \sum_{n=0}^\infty \sum_{j=1}^D \,||n\rangle\rangle_j
\left [ F_n \right ]_j= \sum_{n=0}^\infty \,||n\rangle\rangle\cdot
{ F_n }\ . \label{solpar}
 \ee
In a little bit vague sense it looks like an immediate
hypergeometric-like generalization of eq. (\ref{inser}). Equation
(\ref{reccc}) defines all its coefficients in closed form. They
depend on the ``measure of asymmetry" $a$ and on the unknown
energy $E=-\kappa^2$.

\section{Example }

Our recipe strongly resembles the Hill-determinant method which
proves useful in many (e.g., perturbative \cite{pert})
applications. In the majority of similar applications one must
analyze, first of all, the convergence of infinite series
(\ref{sol}) or (\ref{solpar}). In $x-$representation their
point-wise convergence is basically controlled by the asymptotics
of the coefficients. They are dominated by the purely kinetic
terms which are asymptotically increasing. All the characteristics
of the potential itself (e.g., parity mixing) will play,
necessarily, a secondary role.

The first non-trivial asymmetric potential $V^{[AS]}(x)$ seems
best suited for a more explicit illustration of this role. Its
coefficients $h_j=h_j^{(q)}(p)$ in both the $p=0$ and $p=1$
solutions (\ref{sol}) are easily derived from the respective
recurrences. Choosing the simplest $a=0$ and using the same
abbreviations $a_j$ and $b_j$ as above we have
 \be Q^{(0)}= \left(
\begin{array}{c|cc|cc|cc|c}
0&&&&&&&\\ \hline g&a_1&&&&&&\\ b_0&g&a_2&&&&&\\ \hline
0&b_2&g&a_3&&&&\\ &-g&b_2&g&a_4&&&\\ \hline &&0&b_4&g&a_5&&\\
&&&-g&b_4&g&a_6&\\ \hline &&&&\ddots&\ddots&\ddots&\ddots \ea
\right), \label{jura}
 \ee
  \be Q^{(1)}= \left(
\begin{array}{cc|cc|cc|c}
0&&&&&&\\ g&a_1&&&&&\\ \hline b_1&g&a_2&&&&\\ -g&b_1&g&a_3&&&\\
\hline &0&b_3&g&a_4&&\\ &&-g&b_3&g&a_5&\\ \hline &&& \ddots
&\ddots & \ddots&\ddots \\ \ea \right)\ . \label{jurb}
 \ee
The contribution of the coupling $g$ is clearly separated from the
growing and energy-dependent kinetic terms.

After a return to the general $a \neq 0$ we just have to modify
the values of the matrix elements accordingly. We may preserve the
reduction of bases (\ref{asya}) and (\ref{asyb}) as well as their
$D=2$ partitioning. It is obvious that the exact Jacobi polynomial
solutions may be reproduced in our $D=2$ language. It is an
instructive exercise to show how this reproduction proceeds.
Firstly, the variability of the parameter $a$ {and} of the energy
or momentum $\kappa$ enables us to achieve a complete
disappearance of the two-by-two submatrix $B_K=0$ at an arbitrary
optional $K$. The resulting series (\ref{solpar}) then strictly
terminates and reproduces the known Gauss hypergeometric solution.
Its termination just reflects the factorization of the secular
determinant.

Let us underline that the simpler, ``termination-incompatible"
basis (\ref{party}) with $a=0$ is an analogue of the non-WKB bases
in Section 2.2. Hence, we may fix $a=0$ and recall the same AS
model also as one of the simplest illustrative examples of a
general non-terminating solution.

\subsection{AS oscillator in the $a=0$ representation}

The AS solutions (\ref{sol}) may be split in the two separate sums
with the well defined parity,
\be
|Y^{[AS]}\rangle =|Y^{[AS]}(p)\rangle = |Y^{(even)}(p)\rangle
+|Y^{(odd)}(p)\rangle. \label{old}
 \ee
The first few terms in the even partial sums with $q=1$,
 \ben
|Y^{(even)}(0)\rangle = |\Xi_1\rangle \cdot\, h_0^{(1)}(0) +
|\Xi_5\rangle \cdot\, h_2^{(1)}(0) + |\Xi_9\rangle \cdot\,
h_4^{(1)}(0) + \ldots, \een
\be
|Y^{(even)}(1)\rangle = |\Xi_3\rangle \cdot\, h_1^{(1)}(1) +
|\Xi_7\rangle \cdot\, h_3^{(1)}(1) + |\Xi_{11}\rangle \cdot\,
h_5^{(1)}(1) + \ldots, \label{aold}
 \ee
as well as their odd, $q=0$ counterparts
 \ben |Y^{(odd)}(0)\rangle
= |\Xi_4\rangle \cdot\, h_1^{(0)}(0) + |\Xi_8\rangle \cdot\,
h_3^{(0)}(0) + |\Xi_{12}\rangle \cdot\, h_5^{(0)}(0) + \ldots,
 \een
\be
|Y^{(odd)}(1)\rangle = |\Xi_2\rangle \cdot\, h_0^{(0)}(1) +
|\Xi_6\rangle \cdot\, h_2^{(0)}(1) + |\Xi_{10}\rangle \cdot\,
h_4^{(0)}(1) + \ldots, \label{beold}
 \ee
are easily computed in the recurrent manner,
 \ben h_0^{(1)}(0)=1, \ \
h_1^{(0)}(0)=-g/a_1, \ \ h_2^{(1)}(0)=-b_0/a_2+g^2/(a_1a_2), \ \
\ldots,
 \een
 \be
h_0^{(0)}(1)=1, \ \ h_1^{(1)}(1)=-g/a_1, \ \
h_2^{(0)}(1)=-b_1/a_2+g^2/(a_1a_2), \ \ \ldots\ . \label{cuka}
 \ee
A compact general determinantal formula for these coefficients
also exists \cite{Classif}. It would enable us to re-write eq.
(\ref{old}), i.e.,
 \be
|Y^{[AS]}{(p)}\rangle =
  \sum_{j=0}^\infty \, |\Xi_{\mu(j,p,1)}\rangle \cdot\,
h_{2j+p}^{(1)}(p) +
 \sum_{j=0}^\infty \,
|\Xi_{\mu(j+1-p,p,0)}\rangle \cdot\, h_{2j+1-p}^{(0)}(p)
  \label{Ans0} \label{unusu}
  \ee
in the explicit form if needed. Here, we prefer the recurrent
generation of the doublets of coefficients
 \ben
F_{n+1-p} =F_{n+1-p}(p) =\left( \ba h^{(0)}_{2n+1-p}(p)\\
h^{(1)}_{2n+2-p}(p) \ea \right) , \ \ \ \ \ \ \ p = 0\ {\rm or}\
1, \ \ \ \ \ \  \ n = 0, 1, \ldots \
 \een
as a matrix product,
\be
F_j(p)=\left[-A_j(p)\right]^{-1}B_j(p)\,F_{j-1}(p), \ \ \ \ \ \ \
\ \ \ j = 1,2,\ldots\ . \label{SErec}
 \ee
In our partitioned notation with $D=2$ the solution $|y\rangle$
may be presented as a two-dimensional hypergeometric series since
its matrix coefficients remain surprisingly elementary,
 \ben
\left[-A_j(p)\right]^{-1}= \left(
\begin{array}{cc}
1&0\\ 0&1/a_{2j+p} \ea \right) \left(
\begin{array}{cc}
1&0\\ g&1 \ea \right) \left(
\begin{array}{cc}
1/a_{2j+p-1}&0\\ 0&1 \ea \right).
 \een
As long as $0 > a_1 > a_2 >\ldots$ at any $\kappa > 0$, all our
vectors of coefficients are well defined and unique. Their
initialization is provided by the ``model space" equation
$A_0(p)F_0(p)=0$ which depends on $p$.  At $p=0$, we have the
vanishing scalar $A_0(0 )\equiv 0$ while the exceptional singlet
$F_0(0) =h^{ (1)}_0(0)$ (conveniently put equal to one) is the
norm.  In the parallel two-dimensional initialization at $p=1$,
the first component $ h^{(0)}_0(1)=1$ of $F_0(1)$ is the norm. The
second component must be re-calculated, $h^{(1) }_0(1)=
g\,h_0^{(0)}(1)/(2\kappa+1) $.

We are ready to prove the convergence. Its decisive simplification
occurs in the $j \gg 1$ asymptotic domain. The upper and lower
components of eq. (\ref{SErec}) decouple there in a
$p-$independent manner,
\be
\left[F_j(p)\right]_q \ \left[ \equiv h^{(q)}_{2j+p+q-1}(p)\right]
=
\left[ 1+{1-4q \over 2j}+{\cal O}\left({1 \over j^{2}}\right)
\right] \left[F_{j-1}(p)\right]_q\ , \ \ \ \ \ \ \ \  q = 0\ {\rm
or}\ 1. \label{aSEre}
 \ee
For both our infinite series (\ref{unusu}) the proof is easy at $x
\neq 0$. As long as $\cosh x> 1$ for all the nonzero and real
coordinates $x$, the ordinary geometric criterion together with
the estimate (\ref{aSEre}) implies that our series (\ref{unusu})
are convergent absolutely, i.e., for all the (complex) couplings
$g$ and energies $-\kappa^2$. The same geometric argument extends
the validity of our conclusion to all the complex coordinates
$x+iy$ which lie out of a wiggly bounded domain such that $|\cosh
(x+iy)| =\sqrt{\sinh^2 x + \cos^2 y}\leq 1$ or, in a cruder
approximation, out of the fairly narrow strip with $|x| \leq \ln
(1+\sqrt{2})$ at least.

On the real axis, an indeterminate behaviour of the type $0 \times
\infty$ emerges at the point $x= 0$. This follows from eq.
(\ref{aSEre}) and from the slightly more sophisticated Raabe
criterion. Strictly speaking,  this forces us to work on a
punctured domain of $x \in (-\infty,0 )\bigcup (0,\infty)$ in
principle. As a consequence, logarithmic derivatives of our left
and right Jost solutions have to be matched in the origin. This
task is to be fulfilled numerically. Let us outline its two steps.

\subsection{Generalized parity}

Since our $D=2$ hypergeometric AS series $ \langle x|
Y^{[AS]}{(p)} \rangle \equiv \varphi^{(p)}(g,x,\kappa)$
(\ref{unusu}) satisfy the differential Schr\"{o}dinger equation on
a punctured domain $(-\infty,0) \bigcup (0,\infty)$ only, we
necessarily have to match them in the origin. In the
P\"{o}schl-Teller example of section 2.3 where the non-matrix
Gauss solutions also developed a certain discontinuity in the
origin at a general unphysical energy $E$, the point has easily
been settled  after an account of parity. As long as our
potentials lose their spatial symmetry in general, the parity is
broken and a matching of the two sub-intervals $(-\infty,0)
\bigcup (0,\infty)$ becomes nontrivial.

We have to employ a broader invariance of our model(s) with
respect to the product $\hat{P}$ of parity ${\cal P}$ with the
reflections of couplings $g_j\to -g_j$. The operator (such that
$\hat{P}^2=1$) commutes with our Hamiltonian(s), $H =\hat{P}\, H\,
\hat{P} $. Each physical bound state $\psi(x)$ may be assigned an
even or odd $\hat{P}-$parity, $\hat{P}\,\psi(x) =\pm\psi(x)$.

In a way resembling the parity-breaking systems with ${\cal P T}$
invariance \cite{Bender} the assignment of the $\hat{P}-$parity to
our AS states $\chi(g,x)$ depends on their normalization,
 \ben
 \{ \hat{P} \chi(g,x) = \pm \chi(g,x) \} \ \
\Longrightarrow \ \ \{ \hat{P}\,[ g\cdot \chi(g,x)] = \mp [g\cdot
\chi(g,x)] \} .
 \een
Fortunately, our AS coefficients $h^{(q)}_n(p)=h^{(q)}_n(p,g)$ are
explicitly defined by the triangularized Hamiltonians (\ref{jura})
and (\ref{jurb}) and we immediately notice that
 \ben
h_j^{(q)}(p,-g) = (-1)^{p+q+1}\, h_j^{(q)}(p,g) .
 \een
Both our AS hypergeometric-like series $ \varphi^{(p)}(g,x,\kappa)
= \langle x| Y^{[AS]}{(p)} \rangle$ (\ref{unusu}) behave as
eigenstates of our double-parity operator $\hat{P}$,
 \ben
\hat{P}\,\varphi^{(p)}(g,x,\kappa)= \varphi^{(p)}(-g,-x,\kappa)=
(-1)^p\, \varphi^{(p)}(g,x,\kappa)\ .
 \een
With a pair of some constants ${\cal M} \neq {\cal M}(g)$ and
${\cal N} \neq {\cal N}(g)$ we may postulate that the bound states
read
 \be
\psi^{[AS]}(x)= {\cal M}\, \varphi^{(0)}(g,x,\kappa)+ g\cdot{\cal
N}\, \varphi^{(1)}(g,x,\kappa), \ \ \ \ \ \ \ \ x \neq 0.
\label{nephys}
 \ee
The same (conventionally, even) $\hat{P}$-parity may be assigned
to all our physical solutions since their energy spectrum is
non-degenerate.

\subsection{Match in the origin}

A return to the ordinary spatial parity ${\cal P}$ enables us to
distinguish between the cosine-like (i.e., spatially even) and
sine-like (i.e., spatially odd) components of our generalized
hypergeometric functions (\ref{unusu}),
 \ben c(x,\kappa) ={1 \over
2}[ \varphi^{(0)}(g,x,\kappa) + \varphi^{(0)}(g,-x,\kappa) ], \een
\ben \tilde{s}(x,\kappa) ={1 \over 2}[ \varphi^{(0)}(g,x,\kappa) -
\varphi^{(0)}(g,-x,\kappa) ], \een \ben \tilde{c}(x,\kappa) ={1
\over 2}[ \varphi^{(1)}(g,x,\kappa) + \varphi^{(1)}(g,-x,\kappa)
], \een \ben
 {s}(x,\kappa) ={1 \over 2}[
\varphi^{(1)}(g,x,\kappa) - \varphi^{(1)}(g,-x,\kappa) ].
 \een
The tildas $\tilde{\ }$ marking the asymptotical subdominance are
not too relevant since we dwell in a vicinity of the origin where
$x =\pm \varepsilon \approx 0$.  Wavefunctions must be continuous
there,
 \ben \lim_{\varepsilon \to 0^+}
\psi^{[AS]}_{({\rm physical})}(\varepsilon) = \lim_{\varepsilon
\to 0^+} \psi^{[AS]}_{({\rm physical})}(-\varepsilon).
 \een The
even, cosine-like components of our solutions satisfy such a
requirement identically. In the light of eq. (\ref{nephys}) we are
left with a reduced continuity condition
 \be {\cal M}
\,\tilde{s}(\varepsilon,\kappa) + g\cdot{\cal
N}\,{s}(\varepsilon,\kappa) = 0, \ \ \ \ \ \varepsilon \to 0.
 \label{32a}
 \ee
In the same manner, the continuity of derivatives is required. In
the upper-case notation with abbreviations
 \ben {S}(x,\kappa) ={1 \over 2}[
\partial_x
\varphi^{(0)}(g,x,\kappa) +
\partial_x
\varphi^{(0)}(g,-x,\kappa) ], \een \ben \tilde{C}(x,\kappa) ={1
\over 2}[
\partial_x
\varphi^{(0)}(g,x,\kappa) -
\partial_x
\varphi^{(0)}(g,-x,\kappa) ], \een \ben \tilde{S}(x,\kappa) ={1
\over 2}[
\partial_x
\varphi^{(1)}(g,x,\kappa) +
\partial_x
\varphi^{(1)}(g,-x,\kappa) ], \een \ben {C}(x,\kappa) ={1 \over
2}[
\partial_x
\varphi^{(1)}(g,x,\kappa) -
\partial_x
\varphi^{(1)}(g,-x,\kappa) ]
 \een
this leads to the second reduced matching condition
 \be {\cal M}
\,{S}(\varepsilon,\kappa) + g\cdot{\cal
N}\,\tilde{S}(\varepsilon,\kappa) = 0, \ \ \ \ \ \varepsilon \to
0.
 \label{32b}
 \ee
In the limit $\varepsilon \to 0$ a root $\kappa(\varepsilon)$ of
the two-dimensional secular equation \ben \det \left(
\begin{array}{cc}
\tilde{s}(\varepsilon,\kappa) & {s}(\varepsilon,\kappa)\\
{S}(\varepsilon,\kappa) & \tilde{S}(\varepsilon,\kappa) \ea
\right) = 0 \een will determine the physical energy. Matrix
elements of this secular equation are convergent series in $t =
\cosh^{-2} \varepsilon< 1$, \ben \tilde{s}(\varepsilon,\kappa) =
\sum_{n=1}^\infty \,h^{(0)}_{n}(0,g)\,t^n, , \ \ \ \ \ \
{s}(\varepsilon,\kappa) = \sum_{n=0}^\infty
\,h^{(0)}_{n}(1,g)\,t^n, \een \ben {S}(\varepsilon,\kappa) =
\sum_{n=0}^\infty (\kappa+2n) \,h^{(1)}_{n}(0,g)\,t^n, , \ \ \ \ \
\ \tilde{S}(\varepsilon,\kappa) = \sum_{n=0}^\infty (\kappa+2n+1)
\,h^{(1)}_{n}(1,g)\,t^n. \een Norms $h^{(1)}_{0}(0,g)=
h^{(0)}_{0}(1,g)=1$ are fixed and the higher coefficients carry
the $\kappa-$dependence.  An analogy with the spatially symmetric
P\"{o}schl-Teller example of section 2.3 is fully restored.

\newpage

\section{Summary}

We described a new approach to the Schr\"{o}dinger bound-state
problem with any Rosen-Morse-like multi-term potential
(\ref{RMG}). For all these forces we have shown how

\begin{itemize}

\item
the ordinary differential Schr\"{o}dinger equation for the wave
functions $\psi(x)$ may be reduced to a linear homogeneous
algebraic problem $Q(E)\vec{h}=0$;

\item
an ``inspired" choice of the Lanczos-like (i.e.,
Hamiltonian-dependent) basis makes the related
infinite-dimensional secular determinant vanish identically, $\det
Q(E)=0$;

\item
the very special (viz., lower-triangular) structure of our
quasi-hamiltonian matrices $Q(E)$ reduces the construction of the
separate Taylor-like coefficients $h_n$ in our wave functions
$\psi(x)$ to the mere (partitioned) two-term recurrences.

\end{itemize}

 \noindent
On a characteristic AS example we have illustrated that

\begin{itemize}

\item
all our solutions $\psi(x)$ are convergent and may be understood
as a certain generalization of the Gauss hypergeometric series
(which further degenerates to the Jacobi polynomials at the
physical energies in the solvable cases);

\item
a certain generalized parity symmetry of our forces enables us to
determine Jost solutions which are compatible with {\em both} our
asymptotic boundary conditions;

\item
via our final two-by-two condition (\ref{32a}) + (\ref{32b}), the
values of the remaining two free parameters (viz., energy and
$p-$mixing) in our Jost solutions may (and have to) be tuned to
their necessary continuity and smoothness in the origin.

\end{itemize}

\newpage

\begin{table}
\caption{Shape-invariant potentials on the line \cite{Khare} }
\label{shapeinp}
\begin{center}
\begin{tabular}{||c|c|c|c|c||}
\hline \hline model &$V(x)$
 &
$V(-\infty)$ & $V(\infty)$ & polynomial
 $\psi(x)$  \\ \hline harmonic &
$\omega^2(x+b)^2$ & $\infty$ & $\infty$ &
 Laguerre\\
 Morse &
$a\,e^{-x} + b\,e^{-2x}$ & $\infty$& 0&
  Laguerre \\
  Rosen-Morse II &
${f}/\cosh^{2} x + g \,\tanh  x$
 &
-g&g&
 Jacobi \\
scarf II & $(f+ g \,\sinh  x)/ \cosh^{2} x$
 &
0&0&
 Jacobi \\
\hline
\hline
\end{tabular}
\end{center}
\end{table}

\end{document}